\documentclass[preprint]{aastex}

\slugcomment{Accepted to the Astrophysical Journal}

\shortauthors{Akerlof et al.}
\shorttitle{ROTSE-I Observations of SGR Events}

\begin{document}

\title{Rapid Optical Followup Observations of SGR Events with ROTSE-I}

\author{C. Akerlof\altaffilmark{1}, R. Balsano\altaffilmark{2}, 
S. Barthelmy\altaffilmark{3}, J. Bloch\altaffilmark{2}, 
P. Butterworth\altaffilmark{3}, D. Casperson\altaffilmark{2}, 
T. Cline\altaffilmark{3}, S. Fletcher\altaffilmark{2}, 
G. Gisler\altaffilmark{2}, J. Hills\altaffilmark{2}, R. Kehoe\altaffilmark{1},
B. Lee\altaffilmark{1,4}, S. Marshall\altaffilmark{5}, 
T. McKay\altaffilmark{1}, A. Pawl\altaffilmark{1}, 
W. Priedhorsky\altaffilmark{2}, N. Seldomridge\altaffilmark{2}, 
J. Szymanski\altaffilmark{2}, J. Wren\altaffilmark{2}}

\altaffiltext{1}{University of Michigan, Ann Arbor, Michigan 48109}
\altaffiltext{2}{Los Alamos National Laboratory, Los Alamos, New Mexico 87545}
\altaffiltext{3}{NASA/Goddard Space Flight Center, Greenbelt, MD 20771}
\altaffiltext{4}{Fermi National Accelerator Laboratory, Batavia, Illinois 60510}
\altaffiltext{5}{Lawrence Livermore National Laboratory, Livermore, California 94550}

\begin{abstract}

In order to observe nearly simultaneous emission from Gamma-ray 
Bursts (GRBs), the Robotic Optical Transient Search Experiment 
(ROTSE) receives triggers via the GRB Coordinates Network (GCN).  
Since beginning operations in March, 1998, ROTSE has also taken 
useful data for 10 SGR events: 8 from SGR~1900+14 and 2 from 
SGR~1806-20.  We have searched for new or variable sources in 
the error regions of these SGRs and no optical counterparts were 
observed.  Limits are in the range 
$m_{\mbox{\footnotesize{ROTSE}}} \approx 12.5 - 15.5$ during 
the period $20\ \mbox{seconds}$ to $1~\mbox{hour}$ after the 
observed SGR events.

\end{abstract}


\section{Introduction}
\renewcommand{\thefootnote}{\fnsymbol{footnote}}
\setcounter{footnote}{0}

Soft Gamma-ray Repeaters (SGRs) were separated from classical 
Gamma-Ray Bursts (GRBs) based on the repetition of transient events 
from a single source and their soft spectra 
\citep{mg81,abh+87b,lfk+87}, which are
generally well-fit by an optically thin thermal bremsstrahlung 
model \citep{nhwk91}.  The events tend to be short ($\sim 0.1~\mbox{s}$), 
but on occasion have been observed for hundreds of seconds (e.g. the 
March 5, 1979 \citep{mgi+79} and August 27, 1998 giant flares 
\citep{cmg98,hcm+99,mca+99}).
The positions of SGRs are all near or in supernova 
remnants \citep{cdt+82,kf93,vkfg94,hkw+99,wkp+98} and all have
quiescent X-ray counterparts \citep{mtk+94,hlk+99,rlk94,wkv+99}.
The quiescent X-ray counterparts to SGR~1900+14 \citep{hlk+99}, 
SGR~1806-20 \citep{kds+98} and possibly SGR~1627-41 (\cite{wkv+99} 
but see also \cite{hsl+00}) are X-ray 
pulsars with several second periods.  Furthermore, these pulsars show
spindown periods which imply dipole magnetic fields of $\approx
10^{14}-10^{15}~\mbox{G}$ \citep{kds+98,ksh+98a}.  This evidence
strongly suggests that the sources of SGR events are highly
magnetized, young neutron stars ('magnetars') \citep{dt92a} 
in which magnetically induced 'starquakes' result in the sudden
release of $\gamma$-rays \citep{td95}.

There is currently nothing known about the spectrum of Soft 
Gamma-ray Repeater (SGR) bursts at energies below a few keV.
In this paper we present optical observations of SGR locations
contemporaneous with SGR $\gamma$-ray events.  The Robotic Optical 
Transient Search Experiment (ROTSE) \citep{kab+99} is 
configured to respond to transient events from the Gamma-Ray Burst 
Coordinates Network (GCN) \citep{bbc+95,barthelmy} and is capable of 
rapidly slewing to the coordinates of a transient event such as 
an SGR burst.  Since 
beginning operations in March 1998, the first generation system, 
ROTSE-I, has triggered on 16 SGR events.  For all of these events, 
GCN alerts were based on data from the Burst and Transient Source 
Experiment (BATSE) \citep{fmw+89}.

\section{Observations}

A typical ROTSE-I response to a BATSE-derived GCN trigger consists of 
a series of {\em direct} exposures centered at the trigger coordinates
followed by a series of {\em tiled} exposures with the mount shifted 
by $\pm 8^{\circ}$ (half the total field of view (FOV)) in both right 
ascension and declination to extend ROTSE-I's coverage of the GCN
error box.  The duration of the exposures and the number of times the 
{\em direct}/{\em tile} sequence is repeated depends on the type of 
trigger received from GCN and its typical delay from the onset of 
the trigger event.  Currently, ROTSE-I makes no distinction in its 
response to GRBs and SGRs.  Most GCN triggers arrive $\sim 5\
\mbox{s}$ after the start of a burst, while some are delayed 5-11 
minutes.  ROTSE-I generally begins observations $< 5\ \mbox{s}$ after 
receiving a GCN trigger.  For observations begun within a few minutes
of a burst, image exposure lengths start at 5 s and are extended first to 
25 s and then to 125 s.  In 1999 we reduced the latter two exposure 
lengths to 20 s and 80 s, respectively.

ROTSE-I has responded to 16 SGR triggers.  Of these, useful data were
obtained for 8 SGR~1900+14 bursts and for 2 SGR~1806-20 bursts.
Sample corrected images for both sources are shown in 
Figure~\ref{fig:SGRs}.  For most of these events, the first useful 
images are delayed by over 2 minutes, either because the SGR position 
is only in tiled images\footnote{The known SGR position errors are small 
compared to the ROTSE-I FOV but the rapidly determined positions 
for short, soft spectrum events are often subject to very large 
uncertainties.} or 
because a disproportionate number of SGR triggers were of the delayed 
types.  The exception is BATSE trigger 6809, for
which we have images starting $< 20\ \mbox{s}$ after the
burst and continuing for one hour.  The durations of the SGR bursts were
all $< 1\ \mbox{s}$ with the exception of BATSE trigger 6798 which was 
a series of bursts emitted over a period of 350 s.  In the response to 
this trigger, two ROTSE images follow emission spikes by $< 5\ \mbox{s}$ 
(see Figure~\ref{fig:6798lc}).  ROTSE-I
SGR responses are summarized in Table~\ref{tab:observations}.

\subsection{Data reduction}\label{sec:data_reduction}

All images have been dark-corrected and flat-fielded.  All objects in 
an image are detected using SExtractor \citep{ba96}.  
The object lists are photometrically and astrometrically calibrated 
against the Tycho Reference Catalogue \citep{hkb+98}.  Since ROTSE-I 
uses an unfiltered CCD, the photometry is color-corrected using 
Tycho $B-V$ to produce a ROTSE-I equivalent V-band magnitude, 
$m_{\mbox{\footnotesize{ROTSE}}}$.  This procedure produces 
photometric errors as small as $0.02~\mbox{mag}$ for stars with 
$m_{\mbox{\footnotesize{ROTSE}}} < 12$.  The 
astrometric errors are typically $1.4''$ ($0.1~\mbox{pixel}$).

\subsection{Discussion}

\subsubsection{Locations}\label{sec:locations}

Currently available localizations for SGRs are a small fraction of the
$16^{\circ} \times 16^{\circ}$ ROTSE-I field of view so that only a
small portion of any image must be searched for new or variable
objects.  The adopted search regions for SGR~1900+14 and SGR~1806-20 
are centered on 
$\alpha(2000) =
19^{\mbox{\footnotesize{h}}}7^{\mbox{\footnotesize{m}}}14^{\mbox{\footnotesize{s}}},\  
\delta(2000) = 9^{\circ}20'$ and 
$\alpha(2000) =
18^{\mbox{\footnotesize{h}}}8^{\mbox{\footnotesize{m}}}39^{\mbox{\footnotesize{s}}},\  
\delta(2000) = -20^{\circ}24'40''$ respectively.  In both cases we
use a $2.5'$ (10 pixel) radius search region.  For SGR~1900+14 the
search region is $\approx 100$ times larger than the statistical IPN error 
region \citep{hkc+99} and $10^6$ times larger than the VLA error 
region \citep{fkb99}.   For SGR~1806-20 the search region is 
$\approx 300$ times larger than the IPN error ellipse \citep{hkc+99}.
The error regions are enlarged so that artificially added objects used
to determine our efficiency (see below) do not always fall on the same 
pixels. 

\subsubsection{Limits}

No new or variable objects were detected in any images within the 
specified search regions.  Limits,
shown in Figure~\ref{fig:limits}, were obtained by determining at 
what magnitude our efficiency for detecting artificially added objects 
falls to 50\%.  For SGR1806-20, we found these added objects with 
SExtractor.  For SGR1900+14, we searched by eye for 
the added objects in all images and used SExtractor for a subset of the 
images; a comparison between the methods indicates that agreement 
is within $0.25~\mbox{mag}$.
The limits are calculated in this manner to avoid two potential
problems.  First, the ROTSE-I plate scale is 
$14.4'' \mbox{pixel}^{-1}$ so that distinct
catalog sources may blend into a single object in a ROTSE-I image.  
Second, $m_{\mbox{\footnotesize{ROTSE}}}$ only corresponds to $m_V$ on
average, limiting the usefulness of single object catalog comparisons.

\subsubsection{Extinction}\label{sec:extinction}

Since both SGR~1900+14 and SGR~1806-20 are at low galactic latitudes,
extinction is a large concern.  Unfortunately, no direct measurements
of the extinction to either SGR exist.  However, infrared 
observations covering the IPN localization for SGR~1900+14
provide estimates of $A_V = 15.4 \pm 1.2~\mbox{mag}$ at 
$2.2-6.6~\mbox{kpc}$ and $A_V = 19.1 \pm 1.2~\mbox{mag}$ at 
$12-15~\mbox{kpc}$ \citep{vlh+96}.  To get an alternate estimate for
the extinction to these SGRs, we turn to the X-ray data.

Both SGR~1900+14 and SGR~1806-20 have been detected as X-ray pulsars 
so that the hydrogen column density measured from the X-ray spectra can 
be used to estimate optical extinction via the relation
$A_V = (4.5 \pm 0.3) \times 10^{-22}\ n_h\ \mbox{mag}$ \citep{gor75}. The 
stated error is entirely statistical while the dominant errors are 
likely to be systematic.  Furthermore, this relation ignores the
possible lack of correlation between X-ray absorption and optical
extinction near the SGR itself.  For example, if self-absorption of
X-rays at the source is important, this relation will over-estimate
the optical extinction (see \cite{gor75} for a detailed discussion). 
Conversely, dust local to the SGR would necessitate increasing the
optical extinction calculated by scaling to $n_h$.  Nevertheless, this 
relation will provide us with useful estimates of the optical 
extinction.

SGR~1900+14 was detected as a pulsar by ASCA and spectral fits
gave a hydrogen column density of
$n_h~=~(2.16 \pm 0.07)~\times~10^{22}~\mbox{cm}^{-2}$
and an estimated distance to the SGR of $\sim 5~\mbox{kpc}$
\citep{hlk+99} which agrees with the
distance estimate to the supernova remnant G42.8+0.6 with which the SGR
appears to be associated \citep{vkfg94}.  Thus we obtain
$A_V \approx 10\ \mbox{mag}$ and the equivalent extinctions in R and I
band of $A_R \approx 7\ \mbox{mag}$ and $A_I \approx 5\ \mbox{mag}$
(see e.g. \cite{ccm89}) respectively.  

For SGR~1806-20, 
$n_h \approx 6 \times 10^{22}~\mbox{cm}^{-2}$ \citep{mtk+94,smk+94} 
gives $A_V \approx 27$, $A_R \approx 20$ and $A_I \approx 13$.  Given 
the enormous extinction to SGR~1806-20, we will restrict the remainder 
of the discussion to SGR~1900+14.

As discussed in Section~\ref{sec:data_reduction}, the ROTSE-I system
does not use a standard passband so that an estimate of the
system's spectral response was required to understand the appropriate 
extinction value to adopt.  In Figure~\ref{fig:response}a, the quantum 
efficiency of the Thomson CCD
is shown along with measurements of the transmission of the Canon
telephoto lens.  The lens transmission above $0.65~\mu\mbox{m}$ was 
measured using a Ti-Sapphire laser and errors were
estimated at a few percent.  The lens transmission below 
$0.65~\mu\mbox{m}$ was measured with HeNe and GreNe lasers with errors 
less than 1 percent.  The dashed curve in the figure is a linear 
least-squares fit to the Ti-Sapphire data matched to a line through 
the HeNe and GreNe points.  This simple parameterization is sufficient 
to investigate the effects of extinction on the spectral response of 
the system.  In the right panel of Figure~\ref{fig:response}, the CCD 
quantum efficiency multiplied by the model of the lens transmission is 
shown along with this same curve additionally multiplied by an 
extinction curve with $A_V = 10$.  There are three points to be noted 
about the ROTSE-I spectral response.  First, although we quote an 
equivalent V-band magnitude, $m_{\mbox{\footnotesize{ROTSE}}}$, the 
ROTSE-I system is best color-matched to R-band.  Second, in the
absence of extinction, the response is dominated by the CCD quantum 
efficiency which extends to $\sim 1~\mu\mbox{m}$.  Finally, when the 
extinction is included, the spectral response is more closely matched 
to a narrow I-band filter.  Given these complications, we will 
conservatively adopt $A_R \approx 7\ \mbox{mag}$ applied directly 
to our values of $m_{\mbox{\footnotesize{ROTSE}}}$ for the remainder 
of this paper.

\subsubsection{Physical interpretation of the limits}
\label{sec:interpretation}
With the calculated extinction to SGR~1900+14, we can now address 
how bright an optical transient
would have to be for ROTSE-I to detect it. Accepting the distance of 
$\sim 5\ \mbox{kpc}$ for SGR~1900+14, a typical ROTSE-I limit of 
$m_{\mbox{\footnotesize{ROTSE}}} \approx 14~\mbox{mag}$ 
gives an absolute magnitude limit of 
$M_{\mbox{\footnotesize{ROTSE}}} \approx -1$.  The extinction of 
$A_R \approx 7~\mbox{mag}$ increases this limit to $M_V \approx -8$, 
about the brightness of a nova.

In the absence of firm predictions for the expected optical emission 
from SGRs following soft $\gamma$-ray events, we can only 
investigate constraints on possible sites for emission.  There are two 
constraints with which we will be concerned, the ROTSE-1 magnitude 
limits expressed as a luminosity and the brightness temperature 
limits they imply assuming various emission regions.  We will consider 
three possible sites for optical emission: the magnetar surface, a
region comparable in size to the light cylinder 
(i.e. the radius at which field lines co-rotating
with the neutron star would travel at the speed of light) and a nebula.

For what follows, we compute the luminosity limit implied by
$m_{\mbox{\footnotesize{ROTSE}}} > 14$.  The spectral 
energy density, scaled to an extinction of $A_R = 7\ \mbox{mag}$, is 
$f_{\nu} \le 6 \times 10^{-23}\ 2.5^{7-A_R}\ \mbox{erg}\ 
\mbox{cm}^{-2}\ \mbox{s}^{-1}\ \mbox{Hz}^{-1}$.  This corresponds to
a limit on the isotropic optical luminosity of 
\begin{equation}
L_{\mbox{\footnotesize{opt}}} \le 3 \times 10^{37}\ 2.5^{7-A_R}\ \mbox{erg}\ \mbox{s}^{-1} 
\sim 0.1\ L_{\mbox{\footnotesize{EDD}}}.\label{eq:opt_lum_lim}
\end{equation}

First, we investigate emission from near the magnetar surface.
SGR events with rise and decay times unresolved with $5\ \mbox{ms}$ time
bins imply a source for the $\gamma$-rays $< 1500\ \mbox{km}$ in extent
\citep{knc+87}.  It is reasonable to assume that this emission
arises near the magnetar surface\footnote{The size of a trapped, optically
thick pair-photon plasma responsible for the emission might be at
most $\sim 10-1000$ times larger in area than the neutron star 
(R. Duncan 2000, private communication), which does not change the
conclusion reached here.}.
SGR events are super-Eddington in the X-ray 
($L \sim 10-10^3 \times\ L_{\mbox{\footnotesize{EDD}}}$) so that there 
is certainly sufficient energy released in the system to produce a 
luminosity $L_{\mbox{\footnotesize{opt}}}$
comparable to that in Equation~\ref{eq:opt_lum_lim}.  However, computing the
brightness temperature via $T_b = f_{\nu} c^2/2 \nu^2 k
\Omega_s$\footnote{For a neutron star, $T \sim 1\ \mbox{keV}$ justifies
the Rayleigh-Jeans form for blackbody emission.} with
$\Omega_s$ the solid angle subtended by the source, gives:
\begin{equation}
T_b < 5 \times 10^{9}\ \mbox{keV}\ d_5^2\ 
r_{s,10}^{-2}\label{eq:t_bright}
\end{equation}
where $d_5$ is the distance 
to the SGR is units of $5\ \mbox{kpc}$ and $r_{s,10}$ is its radius in
units of $10\ \mbox{km}$.  This should be compared with a 
typical temperature of $T \sim 25\ \mbox{keV}$ obtained in spectral fits
to burst data or $T \sim 1\ \mbox{keV}$ for the neutron star surface.  
Therefore, the ROTSE-I limits do not
constrain the optical emission from near the magnetar surface.

Next we consider optical emission from a region comparable in size to the
magnetar's light cylinder, $r_{\mbox{\footnotesize{lc}}} = 2.5 \times 10^5\ \mbox{km}$.  
Here the ROTSE-1 limits imply $T_b < 8\ \mbox{keV}$.
At this distance, the magnetar's dipolar field strength, 
$B \sim 10^6\ \mbox{G}$, gives an upper limit of 
$L \sim B^2 c R^2 \approx 10^{43}\ \mbox{ergs}\ \mbox{s}^{-1}$  
to any radiation derived from magnetic field energy.  The optical
limits presented here are considerably lower than this, but we 
lack a plausible mechanism for producing optical emission at this
distance scale.

Finally, we consider emission from a nebula energized by the SGR
activity.  Seven days after the giant flare of August 27, 1998,
\cite{fkb99} derive a lower limit on the size of the radio nebula around 
SGR~1900+14 of $100\ \mu\mbox{as}$.  This can only be used as a rough 
estimate of the size relevant to the ROTSE-1 observations since it 
is a lower limit and it is derived at late times for a nebula which 
is presumed to be expanding.  The implied limit on optical brightness
temperature\footnote{No longer making the Rayleigh-Jeans approximation.} 
at the scale of $r_{\mbox{\footnotesize{neb}}}$ is 
$T_b < 3 \times 10^{-3}\ \mbox{keV}$.  This is well below the limit of
$T_b = 3 \times 10^4\ \mbox{keV}$ for incoherent synchrotron
emission.  The only issue then is whether or not detectable luminosity
could be generated in a nebula surrounding the SGR.

To estimate nebular optical luminosity, we closely follow
\cite{tav94a} using his scalings except where noted.  During a
super-Eddington SGR burst, a relativistic fireball is expected  
which will sweep up material in the medium surrounding the
SGR, producing a shock which will dissipate energy via synchrotron 
radiation.  Scaling to a $\gamma$-ray energy of 
$\sim 10^{40}\ \mbox{ergs}$, the total energy of the burst is 
$E_{b, T} = 10^{40} \epsilon_r^{-1}\ E_{b, 40}\ \mbox{ergs}$ where 
$\epsilon_r$ is the efficiency of conversion of total burst energy
into soft $\gamma$-rays and $E_{b,40}$ is the soft $\gamma$-ray 
energy scaled to $10^{40}\ \mbox{ergs}$.  The electron-positron pairs
in the fireball will be energized 
by the shock and will radiate with a half-power lifetime of 
$\tau_s \sim 5 \times 10^8 \gamma_b^{-1} B^{-2} \sim 5 \times 10^2
\gamma_{b,6}^{-1} B^{-2}$ where $B$ is the magnetic field and 
$\gamma_b$ the Lorentz factor of 
$e^{\pm}$ pairs accelerated in the shock.  We use $B \sim 1\ \mbox{G}$
as estimated from the VLA angular size lower limit \citep{fkb99} and
$\gamma_b \sim 10^6$ as in \cite{tav94a}.  
To obtain the conversion efficiency, $\epsilon_s$, from total burst
energy to high energy nebular emission \cite{tav94a} scales using the 
high energy efficiency of the Crab nebula, $\epsilon_s \sim 10-20 \%$, 
whereas for optical emission $\epsilon_s = 10^{-2} \epsilon_{s,-2}$ is 
a more appropriate estimate (see e.g. \cite{mt77}).
We note that this is still likely to be an over-estimate of the
optical efficiency since the electron density of the optical 
emitting regions in the Crab nebula is $10^3\ \mbox{cm}^{-3}$ 
(see e.g. \cite{mt77}) while SGR~1900+14 is found outside SNR
G42.8+0.6 \citep{vkfg94} where one might expect densities 
$0.1 - 1\ \mbox{cm}^{-3}$ \citep{tc93}.

With the preceding relations, we can parameterize the expected optical 
luminosity from nebular emission as:
\begin{equation}
L \simeq \frac{\epsilon_{s} E_{b, T}}{\tau_{s}} \sim (10^{35}\
\mbox{erg}\ \mbox{s}^{-1}) \epsilon_{s,-2} \epsilon_r^{-1}
\gamma_6^{-1} B^{-2} E_{b, 40}.\label{eq:model_lum}
\end{equation}
The luminosity limit in Equation~\ref{eq:opt_lum_lim} is roughly two orders
of magnitude greater than that estimated for nebular emission in 
Equation~\ref{eq:model_lum} and is therefore not particularly
constraining.  However, this type of emission is the most likely
target for future observations of the type discussed in this paper.

\subsubsection{Comparison to previous work}
\cite{pdh+84} used a high speed photoelectric photometer attached to a 
$50~\mbox{cm}$ telescope to monitor
the position of SGR~0526-66 for 910 hours.  Three optical bursts were
detected, but no $\gamma$-ray events were detected in coincidence with
these events.  Furthermore, there was insufficient information on
backgrounds to unambiguously identify the optical flashes with
SGR~0526-66.  The 3 events peaked around $m_V \sim 9$ for a few
milliseconds.  Assuming that SGR~0526-66 is at a distance of $50\ {\rm
kpc}$ (see e.g. \cite{bkr98}), and correcting for a reddening of
$E(B-V) = 0.37\ \mbox{mag}$ \citep{vblr92}, this corresponds to 
$L \sim 8 \times 10^{38}\ \mbox{erg}\ \mbox{s}^{-1}$ 
at the peak.  Unfortunately, the shortest ROTSE-1 exposures of
$5\ \mbox{s}$ reach a limit $m_{\mbox{\footnotesize{ROTSE}}} \sim
13$ averaged over the exposure length.
Therefore, the luminosity limit in Equation~\ref{eq:opt_lum_lim} for
SGR~1900+14 is a factor of $\sim 20$ too large to detect flashes
similar to those observed by \cite{pdh+84}.

\subsubsection{Future ROTSE observations of SGRs}
With the large extinction in the direction of the known SGRs, 
a campaign specifically designed to
detect SGRs would utilize a rapid-response detector sensitive in the
$1-10~\mu\mbox{m}$ region of the spectrum.  However, ROTSE-I will 
continue to observe SGR triggers since doing so is a simple extension 
of ROTSE's main GRB response program.
Furthermore, the ROTSE collaboration is developing several 
$45~\mbox{cm}$ aperture telescopes which should be $\sim 5\
\mbox{mag}$ more sensitive than ROTSE-I and may be sensitive enough to 
detect optical emission from SGR~1900+14 at 
the level \cite{pdh+84} reported for SGR~0526-66 or the nebular emission
discussed in Section~\ref{sec:interpretation}.

\section{Conclusion}
We have presented the first limits on optical emission
immediately following SGR bursts for 10 events.  Limits on 
optical transient counterparts are in the range 
$m_{\mbox{\footnotesize{ROTSE}}} \approx 
12.5 - 15.5$ during the period $20~\mbox{seconds}$ to $1~\mbox{hour}$ 
after the bursts.  For the 8 events from SGR~1900+14, these limits 
correspond to extinction-corrected absolute magnitudes of $M \sim
-8$.  A more sensitive ROTSE telescope currently being developed may be
sufficiently sensitive to detect optical emission from a flaring
nebula surrounding SGR~1900+14 or emission similar to that possibly 
seen for SGR~0526-66.

We acknowledge useful discussions with Richard Epstein and Tom Vestrand 
and thank Brad Edwards for assistance measuring the Canon lens response.  
We thank Ersin G\"{o}g\"{u}s, Peter Woods and the BATSE team for
providing SGR peak flux and fluence values.  ROTSE is
supported by NASA under {\it SR\&T} grant NAG5-5101, the NSF under
grants AST-9703282 and AST-9970818, the Research Corporation, the
University of Michigan, and the Planetary Society.  Work performed at
LANL is supposed by the DOE under contract W-7405-ENG-36.  Work
performed at LLNL is supported by the DOE under contract
W-7405-ENG-48.

\clearpage
\figcaption[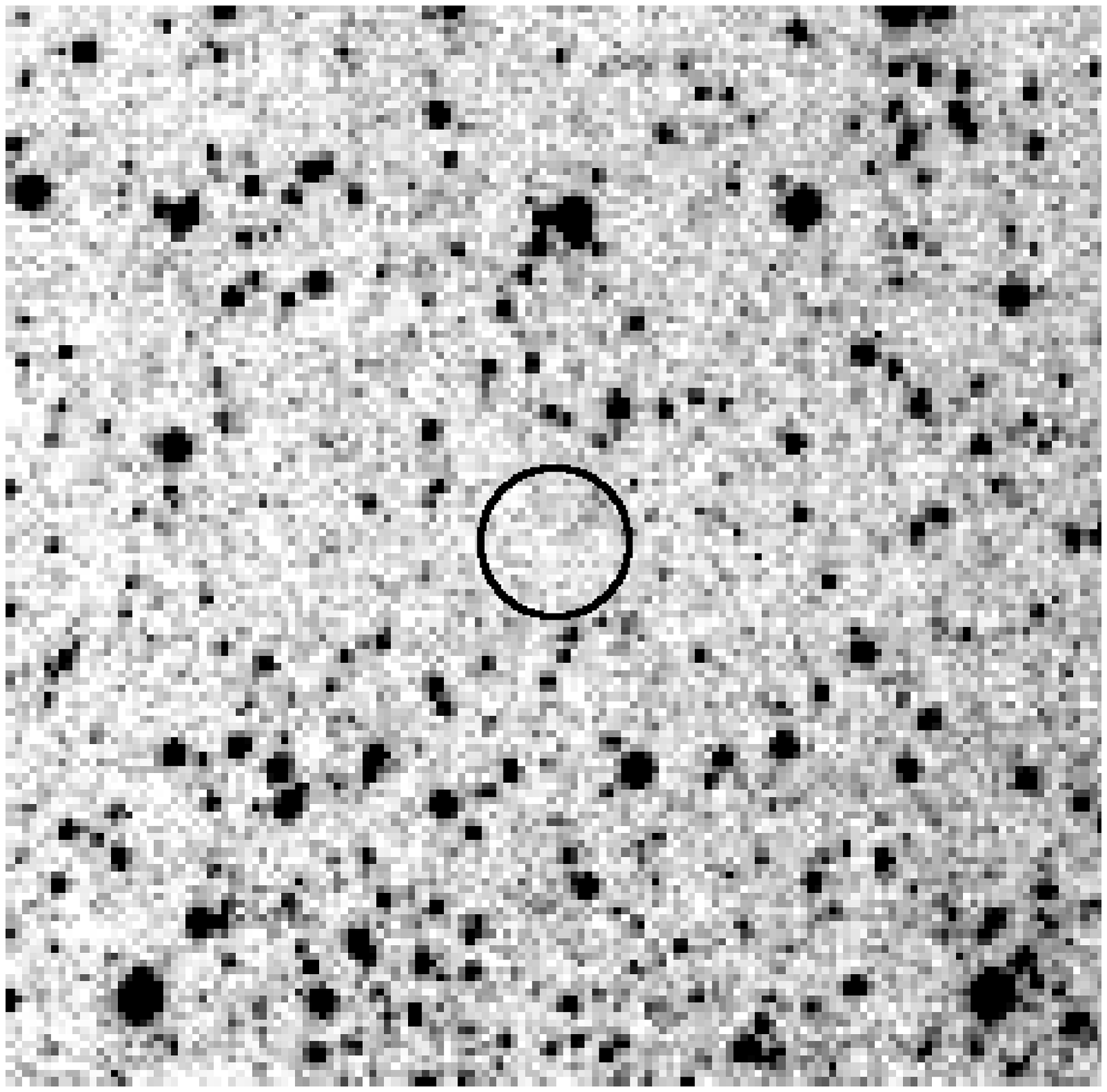,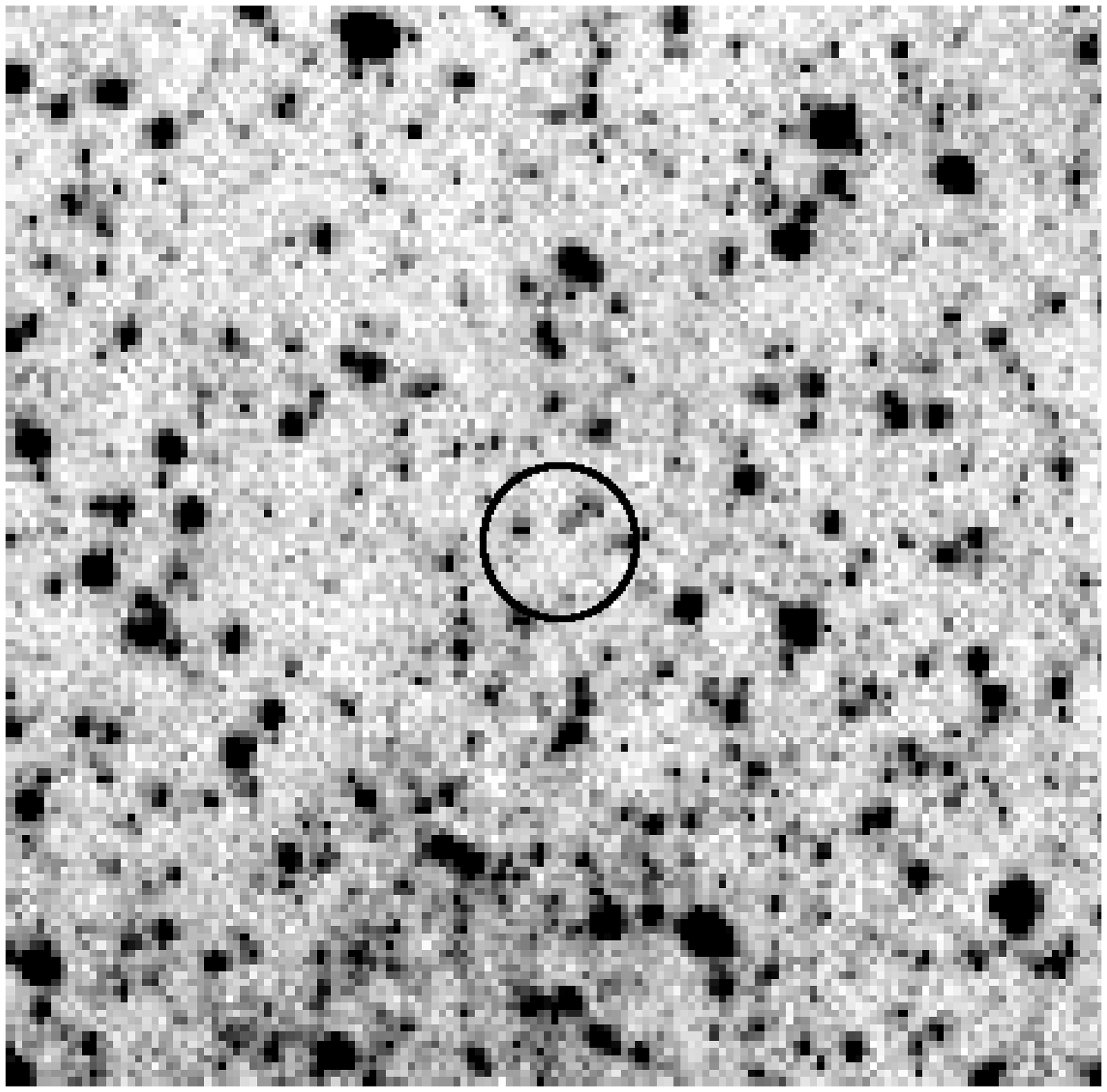]{Sample ROTSE images for (a) SGR~1900+14
	and (b) SGR~1806-20.  Each image is $\approx 50'$
	wide.  The adopted search region, which is $5'$ in diameter,
	is circled.\label{fig:SGRs}}
\figcaption[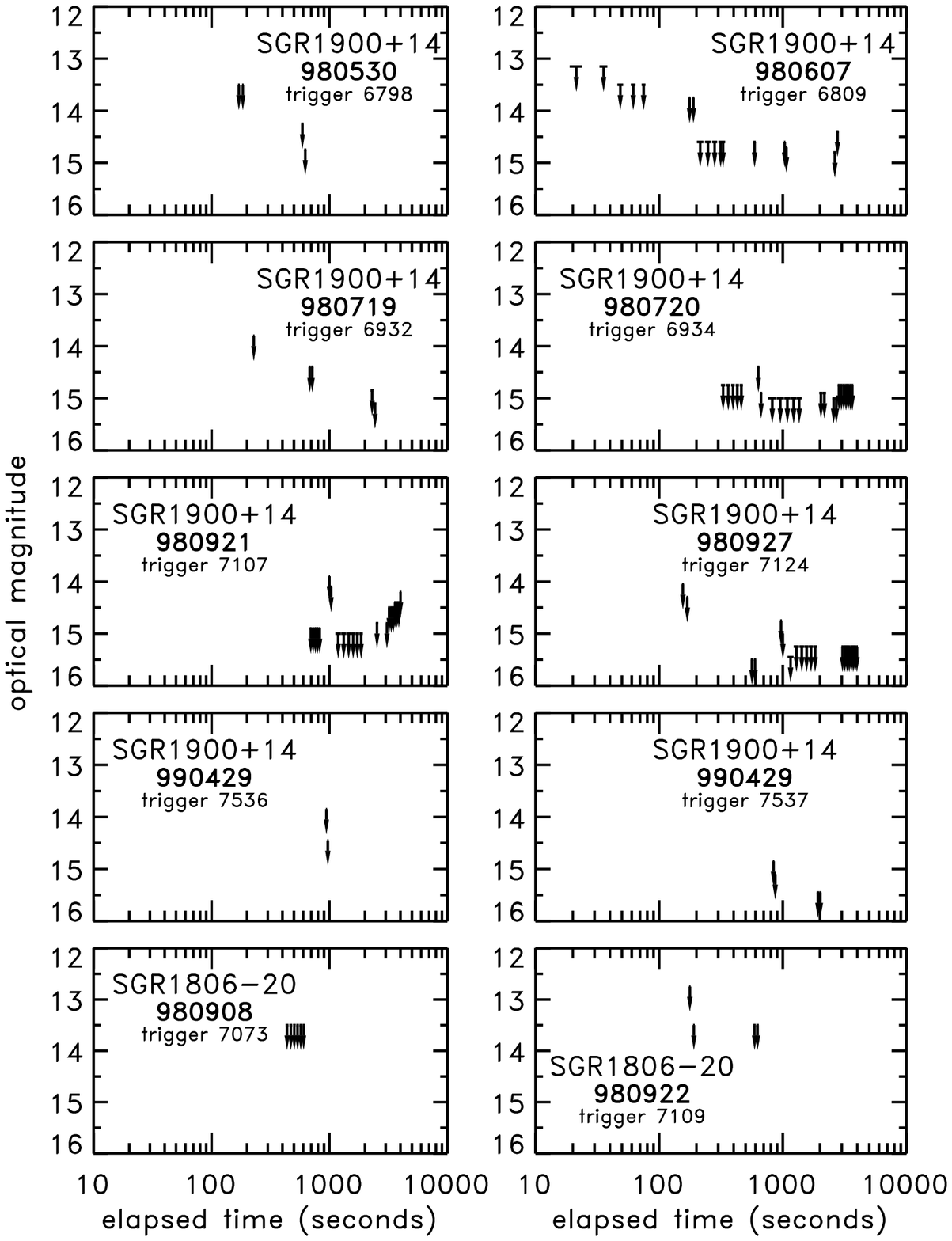]{ROTSE magnitude limits, 
        $m_{\mbox{\footnotesize{ROTSE}}}$, for the 10 triggers with useful 
        data.  Each plot gives the limits, $m_{\mbox{\footnotesize{ROTSE}}}$, 
	as a function of delay since the SGR event start.\label{fig:limits}}
\figcaption[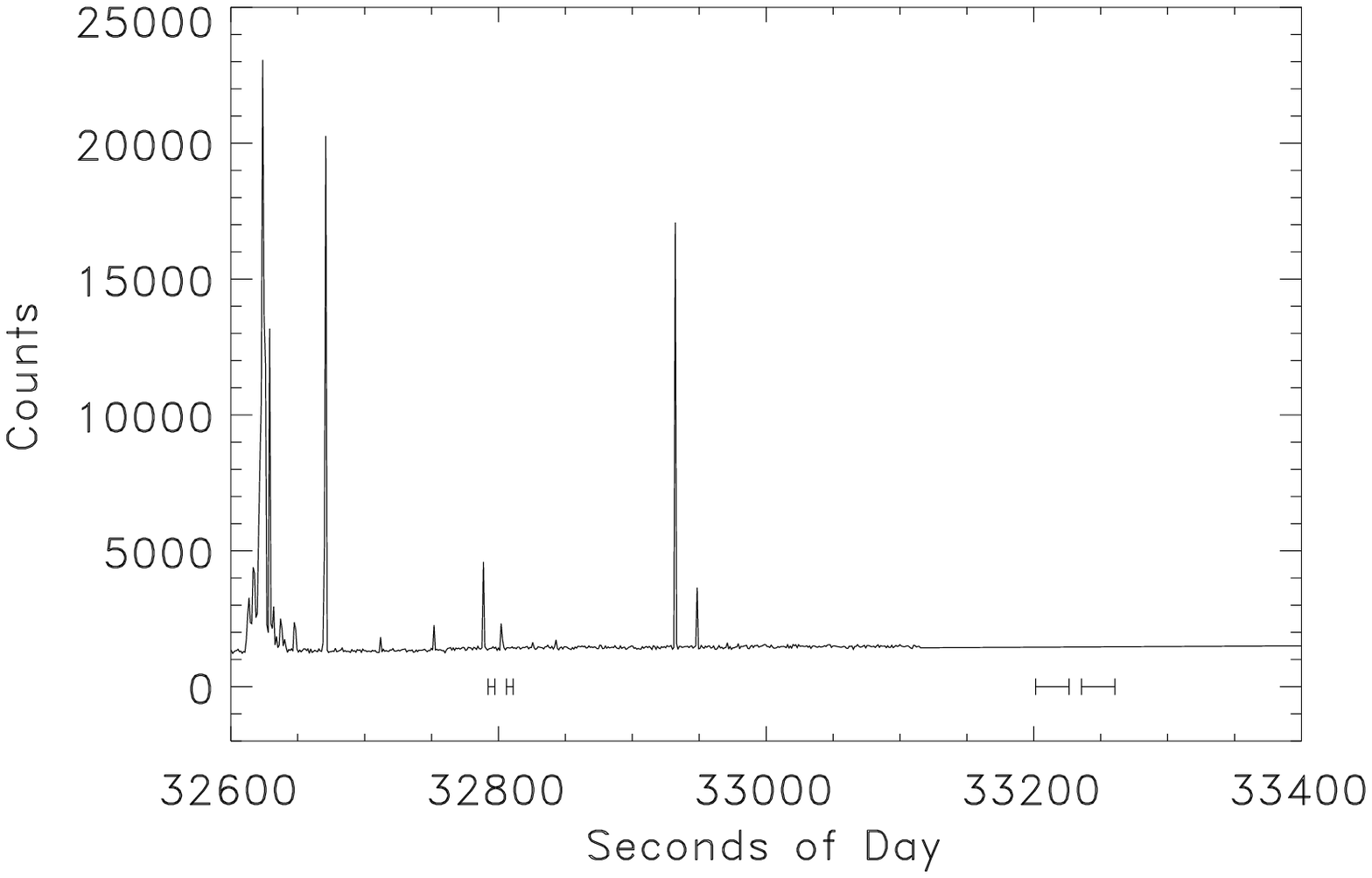]{Raw counts for the SGR~1900+14 event 980530 (BATSE 
	trigger 6798).  The horizontal bars show the times of the ROTSE
	exposures for this event.\label{fig:6798lc}}
\figcaption[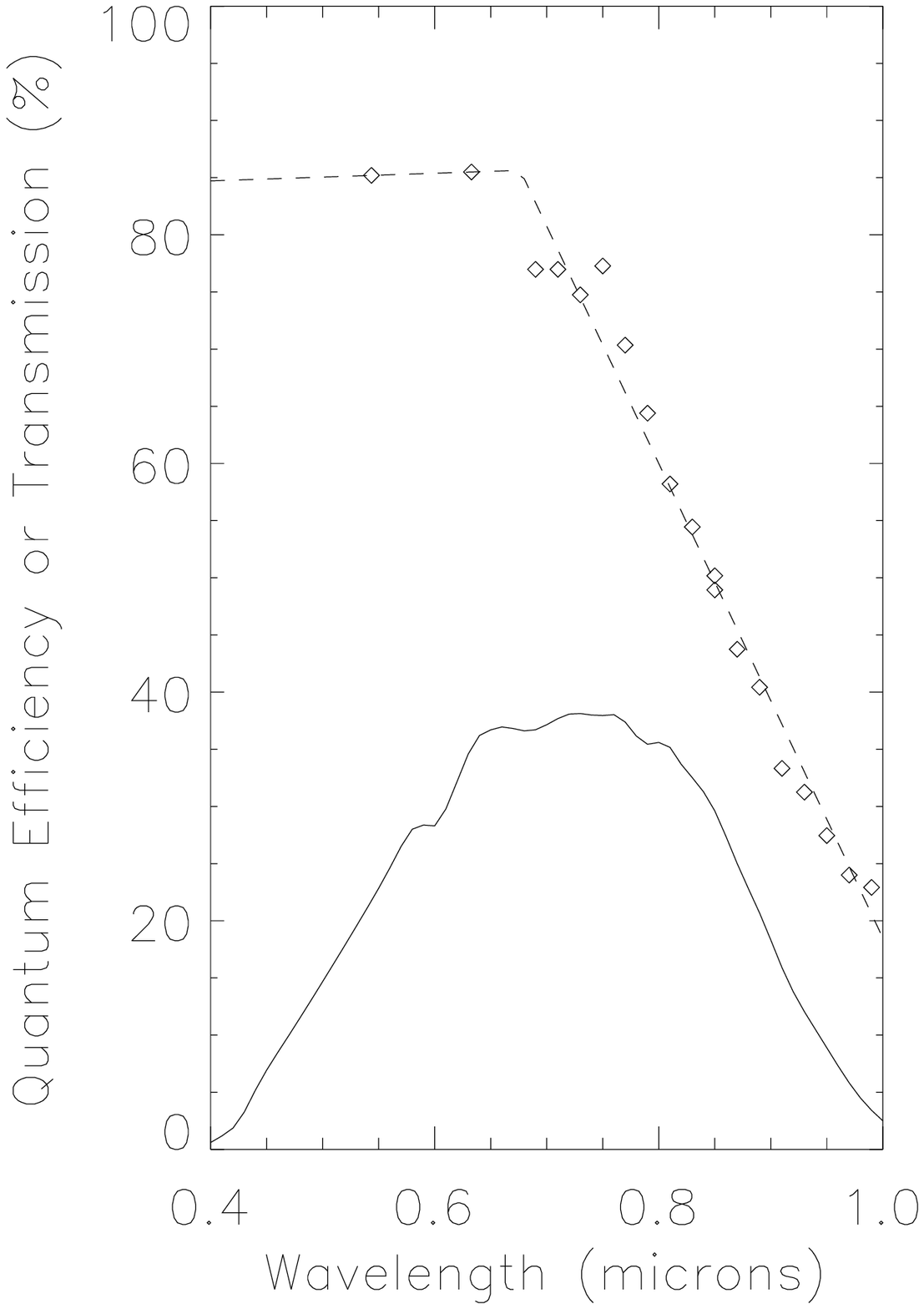,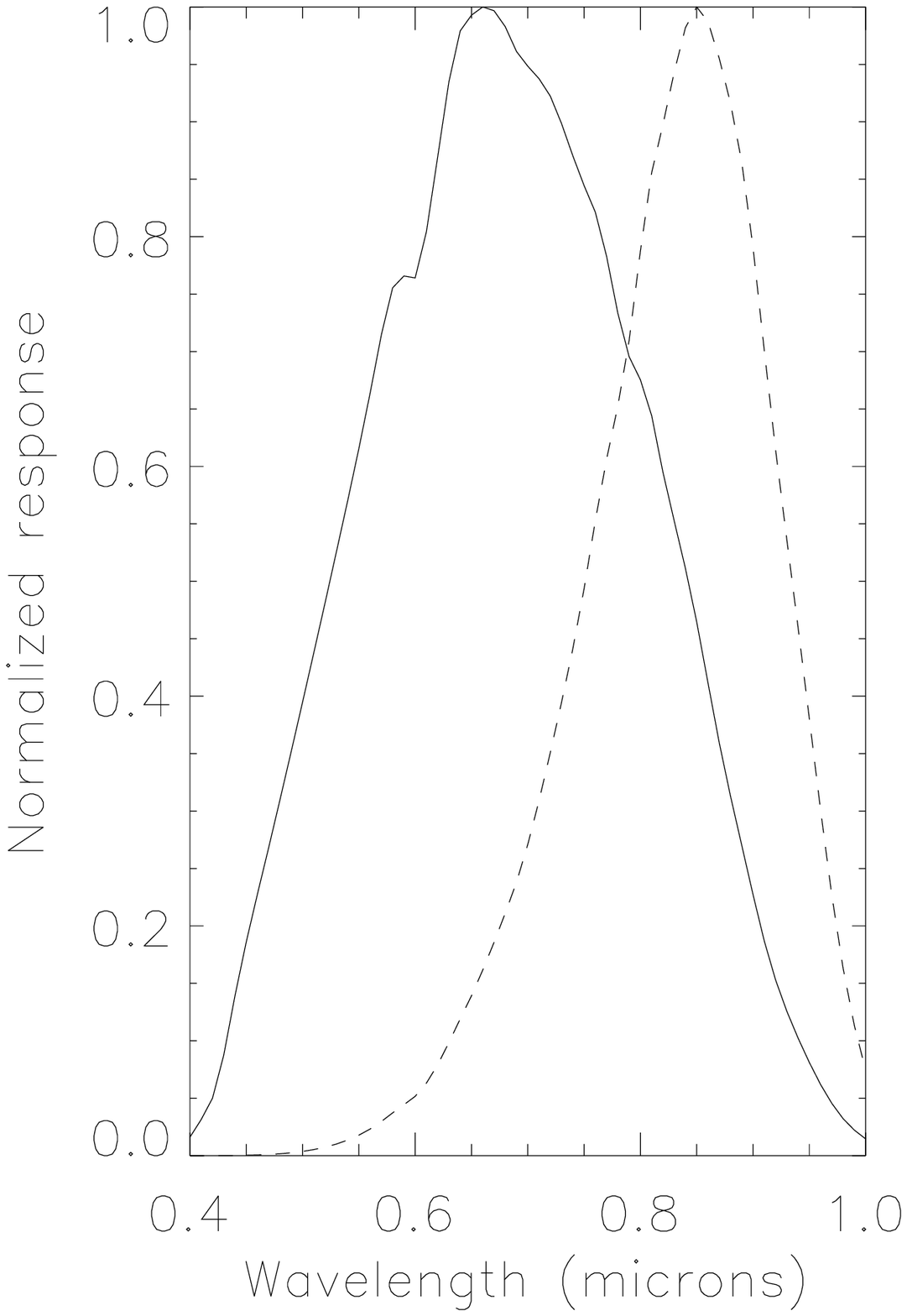]{(a) A typical Thomson THX7899M CCD 
        quantum efficiency (solid line) adapted from \cite{thom95} and the 
	transmission of the Canon FD 200 mm F1.8L lens (see text for a 
	description of the measurements). (b) The normalized CCD+lens 
	response (solid line) and the normalized effective response 
	including an extinction curve with $A_V = 10$ from \cite{ccm89} 
	(dashed line).\label{fig:response}}

\clearpage
\begin{figure}[tb]
\epsscale{0.9}
\plottwo{figure1a.eps}{figure1b.eps}
\end{figure}

\clearpage
\begin{figure}[tp]
  \epsscale{0.95}
  \plotone{figure2.eps}
\end{figure}

\clearpage
\begin{figure}[tb]
  \plotone{figure3.ps}
\end{figure}

\clearpage
\begin{figure}[tb]
\epsscale{1.0}
\plottwo{figure4a.eps}{figure4b.eps}
\end{figure}

\clearpage
\begin{deluxetable}{cccccccl}
\tabletypesize{\scriptsize}
\tablecaption{Summary of ROTSE-I SGR observations.\label{tab:observations}}
\tablewidth{0pt}
\tablehead{
\colhead{SGR}&
\colhead{Date}&\colhead{Trigger}&
\colhead{Peak Flux\tablenotemark{a}}&
\colhead{Fluence\tablenotemark{a}}&
\colhead{$\Delta t$}&
\colhead{$N_{\mbox{im}}$}&
\colhead{Comments\tablenotemark{b}}\\
&&&$(10^{-6}\ \mbox{ergs}\ \mbox{cm}^{-2}\ \mbox{s}^{-1})$&
$(10^{-7}\ \mbox{ergs}\ \mbox{cm}^{-2})$&(s)&&\\
}
\startdata
&980530 & 6798 & 20.4(2) & 354(2) & 168 & 4 & tiles \\
\cline{2-8}
&980607 & 6809 & 14.5(2) & 29.3(2) & 19 & 13 & direct\\
&&&&& 1023 & 4 & tiles \\
\cline{2-8}
&980719 & 6932 & 2.35(5) & 1.52(3) & 226 & 1 & tiles, delayed trigger\\
&&&&& 669 & 4 & tiles\\
\cline{2-8}
1900+14\tablenotemark{c}&980720 & 6934 & 0.71(3) & 0.45(2) & 316 & 23 & direct, delayed trigger\\
\cline{2-8}
&980921 & 7107 & 0.32(3) & 0.24(3) & 682 & 22 & direct, delayed trigger\\
\cline{2-8}
&980927 & 7124 & 5.32(7) & 8.96(8) & 153 & 4 & tiles, delayed trigger\\ 
&&&&& 617 & 16\tablenotemark{d}& direct, delayed trigger\\
\cline{2-8}
&990429 & 7536 & 0.69(3) & 0.54(2) & 933 & 2 & tiles\\
\cline{2-8}
&990429 & 7537 & 0.93(4) & 1.49(5) & 831 & 4 & tiles\\
\tableline
1806-20&980908 & 7073 & 1.44(4) & 1.18(3) & 425 & 6 & direct, first images cloudy\\
\cline{2-8}
&980922 & 7109 & 1.31(4) & 0.98(2) & 174 & 4 & tiles\\
\enddata
\tablenotetext{a}{P. Woods and E. G\"{o}g\"{u}s 2000, private communication}
\tablenotetext{b}{``Direct'' means the SGR was in direct exposures, ``tiles'' 
that it was only in tiles.}
\tablenotetext{c}{For three triggers, a second GCN trigger was
received at a new location.} 
\tablenotetext{d}{This does not include 6 frames in which only half 
of the SGR error region was covered.} 
\tablecomments{Columns specify: SGR source, event date (yymmdd), 
BATSE trigger number, BATSE peak flux 
$(E>25\ \mbox{keV}, 64\ \mbox{ms timescale})$,
BATSE fluence ($E>25\ \mbox{keV}$), delay before first usable image,
number of usable images and comments on the ROTSE data.}
\end{deluxetable}

\end{document}